\documentclass[twocolumn,tighten]{aastex62}
\pdfoutput=1 
\usepackage{amsmath,amstext}
\usepackage[T1]{fontenc}
\usepackage{ae,aecompl}
\usepackage[utf8]{inputenc}
\usepackage{apjfonts} 
\usepackage[figure,figure*]{hypcap}
\usepackage{enumitem}
\usepackage{bm}
\input{hyperlink-year-only-natbib-patch}

\shorttitle{Constraints on DM Microphysics from MW Satellites}
\shortauthors{Nadler, Gluscevic, Boddy, \& Wechsler}

\begin{document}

\title{Constraints on Dark Matter Microphysics from the Milky Way Satellite Population}

\author[0000-0002-1182-3825]{Ethan~O.~Nadler}
\affiliation{Kavli Institute for Particle Astrophysics and Cosmology and Department of Physics, Stanford University, Stanford, CA 94305, USA}

\author[0000-0002-3589-8637]{Vera Gluscevic}
\affiliation{Department of Physics and Astronomy, University of Southern California, Los Angeles, CA 90089-0484, USA}
\affiliation{Joseph Henry Laboratories of Physics, Jadwin Hall, Princeton University, Princeton, NJ 08544, USA}

\author[0000-0003-1928-4667]{Kimberly K.\ Boddy}
\affiliation{Department of Physics \& Astronomy, Johns Hopkins University, Baltimore, MD 21218, USA}

\author[0000-0003-2229-011X]{Risa~H.~Wechsler}
\affiliation{Kavli Institute for Particle Astrophysics and Cosmology and Department of Physics, Stanford University, Stanford, CA 94305, USA}
\affiliation{SLAC National Accelerator Laboratory, Menlo Park, CA 94025, USA}

\correspondingauthor{Ethan~O.~Nadler}
\email{enadler@stanford.edu}

\label{firstpage}

\begin{abstract}
Alternatives to the cold, collisionless dark matter (DM) paradigm in which DM behaves as a collisional fluid generically suppress small-scale structure. Herein we use the observed population of Milky Way (MW) satellite galaxies to constrain the collisional nature of DM, focusing on DM--baryon scattering. We first derive analytic upper limits on the velocity-independent DM--baryon scattering cross section by translating the upper bound on the lowest mass of halos inferred to host satellites into a characteristic cutoff scale in the linear matter power spectrum. We then confirm and improve these results through a detailed probabilistic inference of the MW satellite population that marginalizes over relevant astrophysical uncertainties. This yields $95\%$ confidence upper limits on the DM--baryon scattering cross section of $2\times10^{-29}\ \rm{cm}^2$ ($6\times 10^{-27}\ \rm{cm}^2$) for DM particle masses $m_\chi$ of~$10\ \rm{keV}$ ($10\ \rm{GeV}$); these limits scale as $m_\chi^{1/4}$ for $m_\chi \ll 1\ \rm{GeV}$ and $m_\chi$ for~$m_\chi \gg 1\ \rm{GeV}$. This analysis improves upon cosmological bounds derived from cosmic-microwave-background anisotropy measurements by multiple orders of magnitude over a wide range of DM masses, excluding regions of parameter space previously unexplored by other methods, including direct-detection experiments. Our work reveals a mapping between DM--baryon scattering and other alternative DM models, and we discuss the implications of our results for warm and fuzzy DM scenarios.
\end{abstract}


\section{Introduction}\label{sec:intro}

In the standard cold, collisionless dark matter (CDM) paradigm, structure forms from initial conditions described by a featureless power spectrum of matter perturbations, giving rise to a present-day dark matter (DM) halo mass function that extends uninterrupted down to sub-solar masses \citep{Green0309621,Diemand0501433}.
Non-standard DM scenarios, such as warm DM (WDM; \citealt{Abazajian1710003,Adhikari160204816}) and fuzzy DM (FDM; \citealt{Hu008506,Hui179504}), often involve smoothing of matter perturbations and suppression of structure on small scales.

Herein we focus on the framework inspired by the weakly interacting-massive-particle paradigm, in which DM scatters with Standard Model particles (i.e., baryons), through a velocity-independent contact interaction.
These interactions transfer heat and momentum between the photon--baryon and DM fluids in the early universe, and damp matter perturbations on scales that enter the cosmological horizon while scattering is efficient (e.g., \citealt{Boddy189808}).
The damping scale is set by the interaction strength, and the cumulative effect is largest for the smallest modes, which spend the most time inside the horizon.

As perturbations grow, this distinct fingerprint of DM microphysics is propagated to visible tracers of matter throughout cosmic history as a suppression of small-scale structure relative to CDM.
Currently, the best cosmological limits on DM--baryon scattering come from measurements of damping tails in the cosmic-microwave-background (CMB) temperature and polarization power spectra from \textit{Planck} \citep{Boddy189808,Gluscevic1812108,Xu189710} and the Lyman-$\alpha$ forest flux power spectrum from the Sloan Digital Sky Survey (SDSS; \citealt{Dvorkin890212,Xu189710,Ooba190200826}). Moving beyond probes of quasi-linear cosmological perturbations, the suppression imprinted on the matter power spectrum at early times leads to an underabundance of collapsed objects---notably, low-mass DM halos in our Galactic neighborhood, and the faint galaxies that reside within them.
As low-mass halos arise from matter fluctuations on scales far smaller than those captured by the CMB and other high-redshift probes, population studies of nearby dwarf galaxies could deliver dramatic improvements in sensitivity to DM--baryon interactions.

\begin{figure*}
\hspace{1.275cm}
\includegraphics[scale=0.67375]{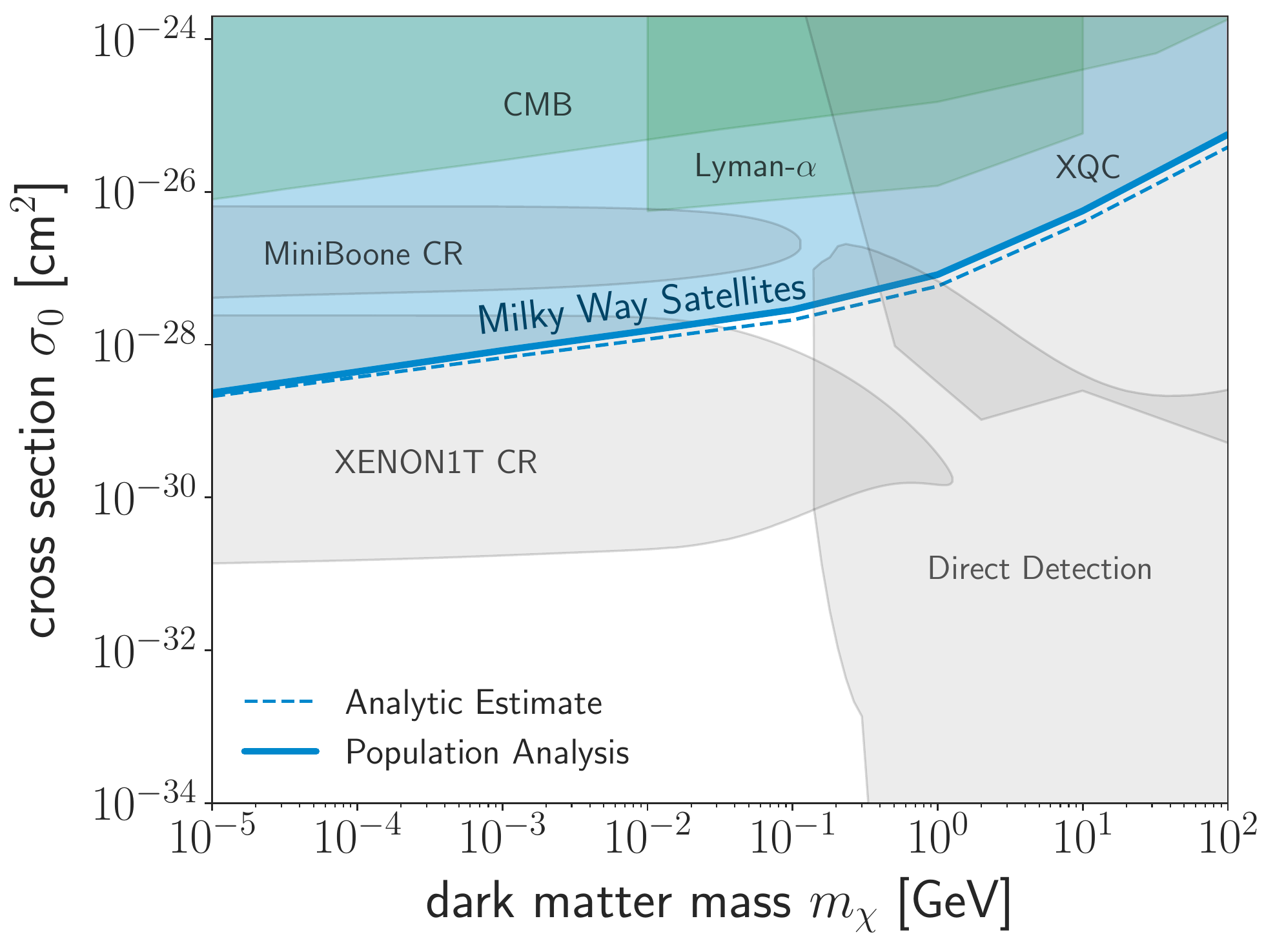}
\caption{
Upper limits on the velocity-independent DM--proton scattering cross section as a function of DM particle mass. The blue shaded region is excluded by the population of classical and SDSS-discovered MW satellites with $95\%$ confidence by our likelihood analysis, which marginalizes over relevant astrophysical uncertainties (Section \ref{sec:analysis}). The dashed line shows analytic upper limit\textcolor{blue}{s} derived from the existence of the lowest-mass halos hosting satellites (Section \ref{sec:analytic}). Green contours show cosmological constraints from the CMB \citep{Boddy189808, Gluscevic1812108} and the Lyman-$\alpha$ forest \citep{Xu189710}. Gray contours show experimental constraints from cosmic-ray scattering \citep{Bringmann181010543}, the X-ray Quantum Calorimeter (XQC; \citealt{Erickek07040794}), and direct-detection experiments including CRESST-III \citep{CRESST}, the CRESST 2017 surface run \citep{Angloher2017}, and XENON1T \citep{Aprile119181301}, as interpreted by \cite{Emken97115047}. Limits from Galactic center gas clouds \citep{Bhoonah180606857,Wadekar190312190} overlap with parts of the XQC, CMB, and Lyman-$\alpha$ contours for $10^{-3}\ \text{GeV} \lesssim m_\chi \lesssim 100\ \rm{GeV}$, and are omitted for clarity.
\label{fig:limit}}
\end{figure*}

In this work, we analyze the observed population of Milky Way (MW) satellite galaxies to place stringent limits on velocity-independent DM--baryon elastic scattering.\footnote{We consider DM--proton interactions, and therefore constrain both the spin-independent and spin-dependent DM--nucleon scattering cross section. Because we neglect helium, our constraints for spin-independent scattering are conservative for DM masses above $\sim$$1\ \rm{GeV}$ \citep{Boddy180800001}.} We present both an analytic derivation of these limits and a rigorous likelihood analysis in which we marginalize over uncertainties related to the connection between galaxies and halos and the impact of baryonic physics on subhalo abundances.
Our population analysis yields~$95\%$ confidence upper limits on the DM--baryon scattering cross section of~$(2 \times10^{-29},8\times 10^{-29},3\times 10^{-28}, 6\times 10^{-27})\ \rm{cm}^2$ for DM particle masses of $(10^{-5}, 10^{-3}, 10^{-1}, 10)\ \rm{GeV}$, improving upon CMB limits by multiple orders of magnitude~(see Figure~\ref{fig:limit}).
This analysis probes unexplored regions of DM parameter space, and it is complementary to direct-detection constraints that rely on the local DM distribution.

This Letter is organized as follows: in Section~\ref{sec:analytic}, we derive analytic limits on DM--baryon scattering from the existence of low-mass halos; in Section \ref{sec:analysis}, we present our likelihood analysis of the MW satellite population and the corresponding DM--baryon scattering limits; in Section~\ref{sec:discussion}, we translate our results into constraints on other non-CDM models; we discuss caveats and conclude in Section~\ref{sec:conclusion}. Throughout, we adopt the best-fit \emph{Planck} cosmology with Hubble parameter~$h=0.6727$, DM density~$\Omega_m h^2 = 0.1199$, and baryon density $\Omega_b h^2 = 0.0222$ \citep{Planck16}, and we set $c=k_B=1$.

\section{Analytic Estimate from Individual Halos}
\label{sec:analytic}

To develop physical intuition for the effects of DM--baryon interactions on the late-time population of DM halos, we first derive an analytic estimate for the mass of the smallest halo allowed to form in an interacting cosmology.\footnote{Note that \cite{Boehm0012504,Boehm0410591} perform similar derivations for DM--photon and DM--neutrino scattering models.}
We then translate upper limits on the minimum halo mass into upper limits on the DM--baryon scattering cross section.
The limits we obtain in this Section do not depend strongly on observational completeness corrections or on galaxy--halo connection modeling, as demonstrated in Section \ref{sec:sigma0}.

\subsection{Minimum Halo Mass in an Interacting Cosmology}
\label{sec:mmin}

In a non-standard cosmology with DM--baryon interactions, linear matter perturbations smaller than a critical length scale~$\lambda_{\rm{crit}}=2\pi/k_{\rm{crit}}$ are substantially suppressed relative to CDM.
As the universe expands, collision damping affects progressively larger scales; when scattering becomes inefficient, the DM and baryon fluids kinetically decouple.
The size of the largest perturbation entirely erased by DM collisions corresponds to the size of the cosmological horizon when the rate of momentum transfer $R_\chi$ between the DM and baryon fluids drops below the Hubble rate $aH$
\begin{equation}
    {aH = R_\chi \left.\right|_{z=z_\textrm{crit}}},
\label{eq:decoupling}
\end{equation}
where $z_\textrm{crit}$ is the corresponding redshift.
To calculate $z_\textrm{crit}$, we use the expression for the velocity-averaged momentum transfer rate for velocity-independent DM--proton scattering \citep{Boddy189808}
\begin{equation}
    R_\chi =  \mathcal{N}_0a \rho_b Y_p\frac{\sigma_0}{m_\chi+m_p}\Big(\frac{T_b}{m_p} + \frac{T_\chi}{m_\chi}\Big)^{1/2},
    \label{eq:Rchi}
\end{equation}
where $\mathcal{N}_0\equiv2^{7/2}/(3\sqrt{\pi})$; $a$ is the scale factor; $\rho_b$ is the baryon energy density; $Y_p$ is the proton mass fraction; $\sigma_0$ is the velocity-independent DM--proton scattering cross section; $m_\chi$ is the DM particle mass; $m_p$ is the proton mass; and $T_b$ and $T_\chi$ are the temperatures of the baryon and DM fluids, respectively.
The term~$\left(T_b/m_p + T_\chi/m_\chi\right)^{1/2}$ is the thermal dispersion of the relative velocity between the DM and baryon fluids.
At early times, the interactions keep the fluids in thermal equilibrium.
In particular, the heat transfer rate $R'_\chi \equiv (m_\chi / (m_\chi + m_p))R_\chi$ exceeds the Hubble rate, driving the DM temperature to that of the photon--baryon fluid, $T_\chi = T_b = T_0 (1+z)$, where $T_0$ is the CMB temperature today.
Thermal decoupling occurs at $z_\textrm{th}$, when the heat transfer rate decreases sufficiently, such that
\begin{equation}
    aH = R'_\chi \left.\right|_{z=z_\textrm{th}}.
    \label{eq:thermal-decoupling}
\end{equation}
Note that $z_\textrm{th}$ occurs deep within the radiation-dominated era, when $H \approx H_0\sqrt{\Omega_\textrm{rad}}z^2$, where $\Omega_\textrm{rad}\approx 10^{-4}$ is the radiation energy density divided by the critical density today,~$\bar\rho$. After thermal decoupling ($z<z_\textrm{th}$), DM cools adiabatically and $T_\chi = T_0 (1+z)^2 / (1+z_\textrm{th})$.

Solving Equation \eqref{eq:thermal-decoupling} for $z_\textrm{th}$, accounting for the thermal evolution of the DM and baryon fluids in Equation \eqref{eq:Rchi}, and making appropriate substitutions in Equation \eqref{eq:decoupling}, we can find $z_\textrm{crit}$ given the parameters $\sigma_0$ and $m_\chi$ that describe our DM--baryon scattering model.
Next, requiring the critical mode to undergo a full oscillation within the cosmological horizon, we compute the corresponding wavenumber $k_{\rm{crit}}$ via
\begin{equation}
    k_{\rm{crit}} = \left. 2 \Big(\frac{1}{aH}\Big)^{-1} \right|_{z=z_\textrm{crit}}.
    \label{eq:kcrit}
\end{equation}
Modes with $k>k_\textrm{crit}$ in the linear matter power spectrum are extremely suppressed (see Figure \ref{fig:pspec}).

The next step is to determine the mass of a collapsed halo arising from perturbation modes that correspond to $k_\textrm{crit}$.
For this purpose, we use the relation
\begin{equation}
M_\textrm{crit} = \frac{4\pi}{3}\rho_m\Big(\frac{\lambda_\textrm{crit}}{2}\Big)^3 = \frac{4\pi}{3}\Omega_m \bar{\rho}\Big(\frac{\pi}{k_\textrm{crit}}\Big)^3.
\label{eq:mcrit}
\end{equation}
Combining with Equation \eqref{eq:kcrit}, this yields
\begin{equation}
M_\textrm{crit}(\sigma_0,m_\chi) \approx \frac{\pi^4}{6}\frac{(\mathcal{N}_0 Y_p\Omega_b)^2 \Omega_m\bar\rho_{\rm{}}^3}{(H_0\sqrt{\Omega_{\rm{rad}}})^5}\frac{T_0}{m_p}\Big(\frac{\sigma_0}{\widetilde{m}_\chi}\Big)^{2},\label{eq:mcrit_approx}
\end{equation}
where $\widetilde{m}_\chi$ asymptotes to $m_\chi$ for $m_\chi\gg m_p$ and to $(m_\chi m_p^3)^{1/4}$ for $m_\chi\ll m_p$; the expression for intermediate DM mass does not have a closed form. Halos that are less massive than~$M_\textrm{crit}\left(\sigma_0,m_\chi\right)$ do not form in a cosmology with DM--baryon interactions because power vanishes on the corresponding scales due to early-time scattering.

\subsection{Limits on the Interaction Cross Section}
\label{sec:sigma0}

If halos are detected down to a minimum mass $M_\textrm{min}$, interpreting an upper limit on $M_\textrm{min}$ as an upper limit on $M_\textrm{crit}$ yields an upper bound on $\sigma_0$, at fixed~$m_\chi$. Thus, detecting halos with $M<M_{\rm{crit}}$ excludes cross sections that correspond to $k_{\rm{crit}}$.

Many independent astrophysical probes testify to the existence of low-mass halos and set upper limits on $M_{\rm{min}}$. For example, substructure detections in strongly lensed systems \citep{Vegetti12013643,Hezaveh160101388} and the dynamical masses of dwarf galaxies obtained from spectroscopy (e.g., \citealt{Simon10074198}) both imply that halos exist down to a mass of $\sim$$10^{9}\ M_{\rm{\odot}}$. Recent studies of the MW satellite population that model the galaxy--halo connection, completeness corrections of observed satellites, and the impact of baryonic physics on galaxy formation and subhalo abundances have pushed the upper bound on $M_{\rm{min}}$ even lower. For example, \cite{Nadler180905542} report $M_\textrm{min}<5.4\times 10^{8}\ M_\odot$ with~$95\%$ confidence using classical and SDSS-discovered MW satellites, and \cite{Jethwa161207834} derived consistent results. These studies do not include ultra-faint satellites discovered in recent years \citep{Bechtol150302584,Drlica-Wagner150803622,Drlica-Wagner160902148,Koposov150302079,Koposov180406430,Laevens150305554,Laevens150707564,Homma160904346}; accounting for these systems will further improve limits on $M_{\rm{min}}$. In addition, upcoming surveys including The Rubin Observatory Legacy Survey of Space and Time (LSST; \citealt{LSST09120201}) are expected to discover even more faint MW satellites, which could lower $M_{\rm{min}}$ by a factor of~$\sim$$5$ compared to current constraints \citep{Drlica-Wagner190201055}.

The observed MW satellite population likely contains several halos that are near the current limit on the minimum halo mass. Thus, we set $M_{\rm{crit}}<5.4\times 10^8\ M_{\rm{\odot}}$ to derive a bound of $k_{\textrm{crit}} > 30\ h\ \textrm{Mpc}^{-1}$.
Using this scale in our analytic prescription yields $\sigma_0 < 2\times 10^{-29}\ \rm{cm}^2$ for a DM particle mass of $10\ \rm{keV}$.
Constraints for other DM particle masses are shown in Figure \ref{fig:limit}.

It is important to note that these limits scale weakly with~$M_\textrm{min}$. In particular, the upper limit on $\sigma_0$ scales as $M_\textrm{crit}^{1/2}$, so increasing $M_\textrm{min}$ by an order of magnitude only weakens the limit on $\sigma_0$ by a factor of~$\sim$$3$. Thus, these constraints do not depend sensitively on the galaxy--halo connection and completeness correction modeling used to derive $M_{\rm{min}}$, and they shift by small amounts if halo masses obtained from strong lensing or spectroscopic measurements are instead chosen as the reference.

\section{Satellite Population Likelihood Analysis}
\label{sec:analysis}

\begin{figure}
\includegraphics[scale=0.43]{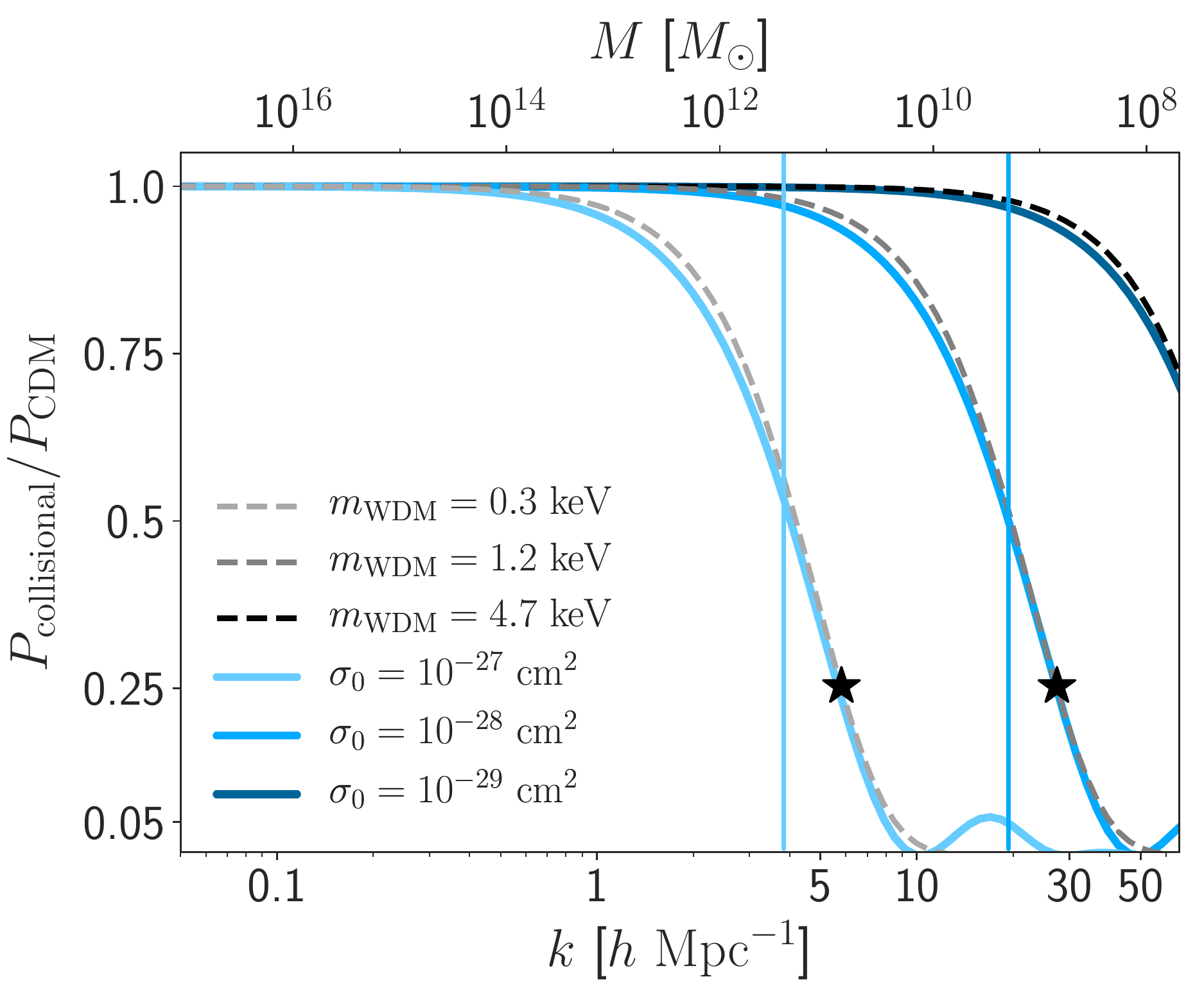}
\caption{Ratio of the linear matter power spectrum in a DM--baryon scattering cosmology to that in CDM (solid lines), for a range of interaction cross sections (for $1\ \rm{MeV}$ DM particles). Dashed lines show the same quantity for WDM models with matching half-mode scales (denoted as black stars). Vertical lines indicate the critical scale discussed in Section \ref{sec:analytic}.
\label{fig:pspec}}
\end{figure}

Halo formation is affected at masses well above $M_{\rm{crit}}$ due to the gradual power suppression caused by DM--baryon interactions (see Figure \ref{fig:pspec}). 
Thus, using a \emph{population} of low-mass halos should yield more stringent limits than the existence of a single low-mass halo. In this section, we therefore perform a probabilistic analysis of the MW satellite population to place more realistic constraints on DM--baryon scattering; our results are shown in Figure \ref{fig:limit}.

In principle, accounting for the detailed effects of DM--baryon scattering on the late-time halo population requires simulations that self-consistently include both the initial linear power suppression described above as well as late-time DM--baryon interactions. However, we find that the power suppression in the DM--baryon scattering case is remarkably similar to that in WDM, as shown in Figure~\ref{fig:pspec}. Moreover, we expect late-time interactions to be a small effect for the interaction model that we consider. Taken together, these facts allow us to use the results of WDM simulations run with nearly identical initial conditions as the DM--baryon scattering model under consideration.

To derive the correspondence between DM--baryon scattering and WDM, we use a modified version of the Boltzmann solver \texttt{CLASS} (described in \citealt{Boddy189808,Boddy180800001,Gluscevic1812108}), which evolves linear cosmological perturbations in the presence of DM--baryon interactions. We generate linear matter power spectra as a function of $\sigma_0$ and~$m_\chi$, and we compare these to WDM power spectra using the transfer function in \cite{Schneider11120330} by matching the \emph{half-mode scale} $k_{\rm{hm}}$, i.e., the wavenumber at which the transfer function~$T(k) = (P_{\rm{collisional}}(k)/P_{\rm{CDM}}(k))^{1/2}$ is equal to~$50\%$. We compute~$k_{\rm{hm}}(\sigma_0,m_\chi)$ numerically using our Boltzmann solver by varying $\sigma_0$ at several values of $m_\chi$.

The correspondence shown in Figure \ref{fig:pspec} allows us to map the suppression in MW subhalo abundances found in WDM simulations to our interacting cosmology. In particular, we use the subhalo mass function from \cite{Lovell13081399}, which is fit to cosmological zoom-in simulations of thermal relic sterile neutrino WDM
\begin{equation}
\frac{\text{d}N}{\text{d}M}\bigg\rvert_{\text{collisional}} = \frac{\text{d}N}{\text{d}M}\bigg\rvert_{\text{CDM}} \Big(1 + \gamma\frac{M_{\rm{hm}}}{M}\Big)^{-\beta},\label{eq:subhalo}
\end{equation}
where $\gamma=2.7$, $\beta=0.99$, $M$ is the peak subhalo virial mass output by the halo finder, and $M_{\rm{hm}}$ is the mass corresponding to $k_{\rm{hm}}$ via a relation equivalent to Equation \eqref{eq:mcrit}.

Next, to forward-model the MW satellite population, we modify the framework presented in \cite{Nadler180905542}. In particular, we supplement high-resolution DM-only simulations of MW-mass host halos \citep{Mao150302637} with a flexible model for the galaxy--halo connection and the impact of both baryonic physics and DM--baryon scattering on subhalo populations. We then fit the luminosity function of classical and SDSS-discovered MW satellites using a Poisson likelihood in bins of satellite luminosity. As in \cite{Nadler180905542}, free parameters in our fit include the slope and scatter in the galaxy--halo connection and the strength of subhalo disruption due to baryonic effects. These are nuisance parameters with large uncertainties that should be marginalized over for the purpose of placing robust limits on DM microphysics. We use the ratio of the collisional-to-CDM subhalo mass functions in Equation \eqref{eq:subhalo} to assign a ``survival probability" to each subhalo in our CDM simulations, following \cite{Jethwa161207834}. 
Thus, the final free parameter in our fit is $M_{\rm{hm}}$, and we obtain a marginalized posterior distribution~$P(M_{\rm{hm}})$ using a flat prior on $\log(M_{\rm{hm}})$. 
To be conservative, we assume that \emph{all} subhalos host galaxies, even though the galaxy occupation fraction is likely low and mass-dependent in this regime (e.g., \citealt{Sawala14066362,Fitts180106187}). A non-trivial occupation fraction could force lighter subhalos to host observed satellites, further strengthening our constraints. However, we find that marginalizing over a step-function galaxy formation threshold does not significantly affect our results.

Finally, we map $P(M_{\rm{hm}})$ to $P(\sigma_0| m_\chi)$ using our half-mode scale calculation. We find $M_{\rm{hm}}<3.1\times10^{8}\ M_{\rm{\odot}}$ (corresponding to $k_{\textrm{hm}} > 36\ h\ \textrm{Mpc}^{-1}$) with $95\%$ confidence, yielding upper limits on $\sigma_0$ of $(2 \times10^{-29},8\times 10^{-29},3\times 10^{-28}, 6\times 10^{-27})\ \rm{cm}^2$ for DM particle masses of $(10^{-5}, 10^{-3}, 10^{-1}, 10)\ \rm{GeV}$, as shown in Figure \ref{fig:limit}. These limits improve upon CMB constraints by multiple of magnitude for $m_\chi\lesssim 1\ \rm{GeV}$, and can be extrapolated to higher masses. However, for $m_\chi\lesssim 10\ \rm{keV}$, relativistic effects become important. Moreover, in the following section we show that thermal DM lighter than $\sim$$3\ \rm{keV}$ is ruled out at all cross sections.

\section{Implications for Other Non-CDM Models}
\label{sec:discussion}

The methods developed in this work apply to \emph{any} model in which DM behaves similarly to a collisional fluid on small scales.
For example, we can translate our $M_{\rm{hm}}$ constraint to a lower limit on WDM mass by combining the transfer function derived in \cite{Schneider11120330} for a thermal relic sterile neutrino of mass~$m_{\rm{WDM}}$ with the half-mode mass equivalent of Equation \eqref{eq:mcrit}, which yields
\begin{equation}
    m_{\rm{WDM}} = 2.32\ \Big(\frac{\Omega_m}{0.25}\Big)^{0.4}\Big(\frac{h}{0.7}\Big)^{0.8}\Big(\frac{M_{\rm{hm}}}{10^{9} M_{\rm{\odot}}}\Big)^{-0.3}\ \rm{keV}.\label{eq:wdm}
\end{equation}
We find $m_{\rm{WDM}}>3.26\ \rm{keV}$ with $95\%$ confidence, which is consistent with previous MW satellite results \citep{Jethwa161207834} and competitive with Lyman-$\alpha$ forest constraints \citep{Viel1308804,Irsic179602}. 

Our results also constrain FDM models, in which ultra-light axions comprise DM and small-scale structure is suppressed due to quantum interference effects \citep{Hu008506,Hui179504}. 
We translate our~$M_{\rm{hm}}$ constraint into a lower limit on FDM mass $m_\phi$ using the $m_{\rm{WDM}}$--$m_\phi$ relation from \cite{Armengaud170309126} in Equation \eqref{eq:wdm}, which gives
\begin{equation}
    m_{\phi} = 1.3\times 10^{-21}\  \Big(\frac{\Omega_m}{0.25}\Big)^{0.95}\Big(\frac{h}{0.7}\Big)^{1.9}\Big(\frac{M_{\rm{hm}}}{10^{9} M_{\rm{\odot}}}\Big)^{-0.71}\ \rm{eV}.
\end{equation}
We find $m_\phi > 2.9\times 10^{-21}\ \rm{eV}$, which is again competitive with Lyman-$\alpha$ forest constraints \citep{Irsic1711903}. We note that the high-redshift galaxy luminosity function and the epoch of reionization provide WDM and FDM constraints that are complementary to---but currently weaker than---our limits (e.g., \citealt{Corasaniti161105892}).

We expect that a re-analysis of the MW satellite population using our method will yield even more stringent limits for velocity-dependent DM--baryon scattering, i.e., for any scattering model in which the cross section scales as the relative particle velocity to a positive power. Such interactions arise in DM effective theory (e.g., \citealt{Boddy189808}), and we leave an investigation of this scenario to future work.

\section{Conclusions and Discussion}
\label{sec:conclusion}

Small-scale tracers of the matter distribution in the universe provide insights into DM microphysics that are complementary to other probes.
In this work, we report stringent upper bounds on DM--baryon scattering for a velocity-independent contact interaction by analyzing the population of classical and SDSS-discovered MW satellite galaxies. 
Based on the consistency of the satellite population with CDM expectations down to a halo mass scale of~$\sim$$10^{8}\ M_{\rm{\odot}}$, we place robust limits on a range of DM interaction cross sections and particle masses.
Our analysis closes unexplored gaps in DM--baryon scattering parameter space, improving upon previous cosmological bounds by several orders of magnitude, and it is not subject to uncertainties in the local DM distribution or cosmic-ray propagation.

In our likelihood analysis, we have accounted for several astrophysical uncertainties, including the slope and scatter of the low-mass galaxy--halo relation and the impact of baryonic physics on subhalo abundances.
We have made the conservative assumption that \emph{all} subhalos host galaxies, though marginalizing over a step-function galaxy formation threshold does not affect our results. Moreover, \cite{Nadler180905542} verified that potential spurious numerical effects in our simulations (e.g., artificial subhalo disruption) do not alter the inferred minimum halo mass.

There are several possible caveats to our analysis that we leave for future work. First, we have not investigated how our results depend on cosmological parameters. However, we expect uncertainties on cosmological parameters to play a minimal role; for example, Equation \eqref{eq:mcrit_approx} implies that limits on $\sigma_0$ scale as $\Omega_m^{-1/2}$. Future analyses that marginalize over cosmological parameters are needed to confirm this weak dependence.
Next, we relied on simulations with a narrow range of host halo mass, even though the uncertainty on the mass of MW is relatively large (e.g., \citealt{Deason180810456}).
We do not expect this uncertainty to affect our results significantly, as $M_{\rm{min}}$ scales linearly with host mass and our limits depend weakly on $M_{\rm{min}}$.
To further improve our constraints, it will be necessary to understand how a mass-dependent galaxy occupation fraction and baryonic effects beyond enhanced subhalo disruption affect satellite populations in detail.

This work demonstrates that the MW satellite population places competitive constraints on a class of non-CDM models. Moreover, it paves the way for joint probabilistic analyses of forthcoming small-scale structure datasets.
For example, upcoming imaging surveys such as LSST will facilitate systematic searches for density gaps in Galactic stellar streams, which potentially trace even lower-mass subhalos than those inferred from satellites~(e.g., \citealt{Bonaca181103631}). 
In addition, gravitational lensing measurements with the Atacama Large
Millimeter/submillimeter Array will map out low-mass DM substructure in a complementary manner to faint galaxies (e.g., \citealt{Hezaveh14032720}).
Finally, future spectroscopic surveys like DESI \citep{DESI161100036} will provide improved Lyman-$\alpha$ forest measurements, allowing for important high-redshift consistency tests of local small-scale structure constraints. 
Distinct systematic and modeling uncertainties accompany each of these DM probes, and joint likelihood analyses of all available datasets---enabled by approaches similar to ours---will be crucial in order to study the nature of DM in the era of next-generation surveys.
\section*{Acknowledgements}

We are very grateful to Alex Drlica-Wagner for assisting with aspects of this work. We thank Arka Banerjee, Keith Bechtol, Yao-Yuan Mao, Annika Peter, and particularly Francis-Yan Cyr-Racine for comments on the manuscript. We thank the LSST DM Group for feedback at workshops supported by the LSSTC Enabling Science program (grant No.\ 2017-11). V.G.\ acknowledges the hospitality of KIPAC, where this work began. K.B.\ acknowledges the support of a Johns Hopkins Provost’s Postdoctoral Fellowship. This research was supported by the NSF Graduate Research Fellowship received by E.O.N.\ (grant No.\ NSF DGE-1656518) and the US DOE contract to SLAC (grant No.\ DE-AC02-76SF00515).  This research utilized computational resources at SLAC National Accelerator Laboratory.

\bibliographystyle{yahapj}
\bibliography{references}

\end{document}